\documentclass[aps,prl,reprint,superscriptaddress,longbibliography,showkeys,floatfix]{revtex4-2}
\usepackage[utf8]{inputenc}
\usepackage[T1]{fontenc}
\usepackage{lmodern}            % Clean, modern Latin Modern fonts
\usepackage{amsmath,amssymb,bm}
\usepackage{mathtools}
\usepackage{graphicx}
\usepackage{microtype}
\usepackage{siunitx}            % \SI{}{} for units
\usepackage{xcolor}
\usepackage{placeins}   % optional
\usepackage{graphicx}
\usepackage{placeins}  % optional: \FloatBarrier to flush earlier floats
\usepackage{capt-of}   % for \captionof{figure}{...}
\usepackage{xcolor}
\usepackage{tikz}
\usetikzlibrary{decorations.pathmorphing}
\usepackage{placeins} % for \FloatBarrier
\usepackage{comment}
\usepackage[colorlinks=true,allcolors=blue!50!black,linktocpage]{hyperref}
\usepackage{latexsym}

\usepackage{xcolor}
\usepackage{tikz}
\usepackage[normalem]{ulem}
\usepackage{xcolor}
\usepackage{graphicx}
\usetikzlibrary{decorations.pathmorphing}
\newcommand{\oldwave}[2][blue]{%
	\bgroup
	\color{black}%
	\markoverwith{%
		\textcolor{#1}{%
			\raisebox{0.15ex}{%
				\scalebox{.6}[1.7]{$\sim$}%
			}%
		}%
	}%
	\ULon{#2}%
}

\begin{document}
\title{Stochastic model of elongation-shortening-switching kinetics of a polymeric filament:\\ Critical slowing down of first-passage and finite-size regularization}
\author{Manvendra Singh}
\affiliation{Department of Physics, The LNM Institute Of Information Technology, Rupa Ki Nangal, Post-Sumel, via - Jamdoli, Jaipur 302031, Rajasthan, India}
\author{Ashok Garai}
\email{corresponding author: ashok.garai@lnmiit.ac.in}
\affiliation{Department of Physics, The LNM Institute Of Information Technology, Rupa Ki Nangal, Post-Sumel, via - Jamdoli, Jaipur 302031, Rajasthan, India}
\author{Debashish Chowdhury}
\email{On leave from DIT University, Dehradun; present address: C-108, Pacific Hills, Mussoorie Diversion, Dehradun.}
\affiliation{Department of Physics, DIT University,\\ Mussoorie Diversion Road, Dehradun 248009, Uttarakhand, India. }

\date{\today}

\begin{abstract}
Several filamentous biopolymers exhibit unusual polymerization-depolymerization kinetics where continued elongation (shortening) is possible only in an `active' (`inactive') state while occasional random transition (switching) between the active and inactive states takes place. Here, we present a minimal stochastic model capturing only the key features of random elongation-shortening-switching kinetics of a generic filament. The structure and/or function of the filament ends when its length hits a critical value during its shortening; the time to reach this critical length can be formulated as a first-passage time (FPT) and for its calculation the critical length is treated as an absorbing boundary. Moreover, the lengths of the filaments that inspire our model are constrained to a finite maximum value $\mathcal{N}$; therefore, in our model we impose a reflecting boundary beyond which the filament cannot grow.  We determined the distribution of the FPT and derived its first two moments. We identify a critical point where the drift of the tip of the filament vanishes making its movement purely diffusive; at this critical point, in the limit $\mathcal{N} \to \infty$, the fluctuations in the FPT diverges and MFPT becomes infinite indicating `critical slowing down'. Our analytical results will serve as the testing ground for the accuracy of the approximate theoretical results of more realistic models for polymerization-depolymerization kinetics of specific bio-filaments.
\end{abstract}

\maketitle

\section{Introduction}
 Competing processes of continued elongation and continued shortening, interspersed by occasional switching between these two phases, is a common feature of several bio-filaments that are essentially linear polymers. The most familiar example of such a system in microtubule (MT) where its polymerization to depolymerization, caused by the loss of a GTP cap, is called `catastrophe' whereas the opposite transition is called `rescue' \cite{brouhard15,gardner13,Buxton2010,Nelson2026}. Unless rescued, a shortening MT eventually disappears when its length vanishes. The time interval between the nucleation  of a MT to its eventual return to the site of its nucleation defines its lifetime which is a random variable because of the intrinsic stochasticity of the polymerization, depolymerization as well as catastrophe and rescue. 
Similarly, Telomeres are repetitive DNA sequences at the ends of linear chromosomes. Because conventional DNA polymerases cannot fully replicate the very ends of linear DNA, telomeres undergo progressive shortening with each cell division \cite{shay00}. When a telomere shortens to a critical length it can trigger replicative senescence. Telomerase is a supra-molecular complex whose key enzymatic activity is to elongate the telomere thereby countering the shortening of the telomere during normal DNA replication. The time needed for a telomere to shorten to a critical length that triggers replicative senescence can be regarded as the lifetime of the replication-competent state of the corresponding gene.

The lifetime of a MT and the lifetime of the replication-competent state of a gene can both be regarded as a {\it first-passage time} (FPT) \cite{Redner2001}. Capturing some of the key features of the shortening-elongation kinetics of the type mentioned above we develop a generic stochastic model for filaments and obtain the exact distribution of the FPT. Because of the finite size of a cell MTs cannot grow beyond a maximum length ${\ell}=\mathcal{N}$. Similarly, in a non-cancerous cell the telomere cannot be longer than a maximum value ${\ell} = \mathcal{N}$. Using the backward master equations \cite{schuster2023} for this model to derive the corresponding equations for the survival probabilities, and assuming the existence of a quasi-stationary distributions of the lengths of the filaments, we derive analytical expressions for the distributions of the FPT under a reflecting boundary condition at ${\ell}=\mathcal{N}$. Moreover, expressions for the mean and variance of the FPT distribution shows the existence of a `critical point' where, because of the vanishing drift velocity of the tip of the filament its kinetics becomes purely diffusive and, in the limit $\mathcal{N} \to \infty$., the fluctuations in the FPT diverge and the divergence of the MFPT implies a `critical slowing down'. \\

Fist-passage times for the capture of the growing tip of a polymeric filament like MT by a binding partner has been studied extensively for almost three decades \cite{holy94} primarily in the context of `search-and-capture' phenomena \cite{heald15,gopalakrishnan11,mulder12,chowdhury19} which has been examined even from the broader perspective where biological processes are viewed as `exploratory dynamics' \cite{kondev25}. However, in the present study we investigate the first-passage of the distal tip of the polymeric filament to its own `base' where, effectively its other tip. Although results for this latter phenomenon, to our knowledge, was reported many years ago \cite{bicout97}, corrections to the results arising from the finite size of the geometry are being reported for the first time in this paper. Microtubules or telomeres serve only as motivating examples, but the theory is not restricted to these two systems alone. The generic one-dimensional stochastic kinetic model can be applied to any such filament that exhibits random switching between the elongating and shortening states, and whose lifetime is defined by the point at which it reaches a pre-fixed minimum length threshold.

%%%%%%%%%%%%%%%%%%%%
\section{Model and master equation}
\begin{figure}[t]
	\centering
	\includegraphics[width=1\linewidth]{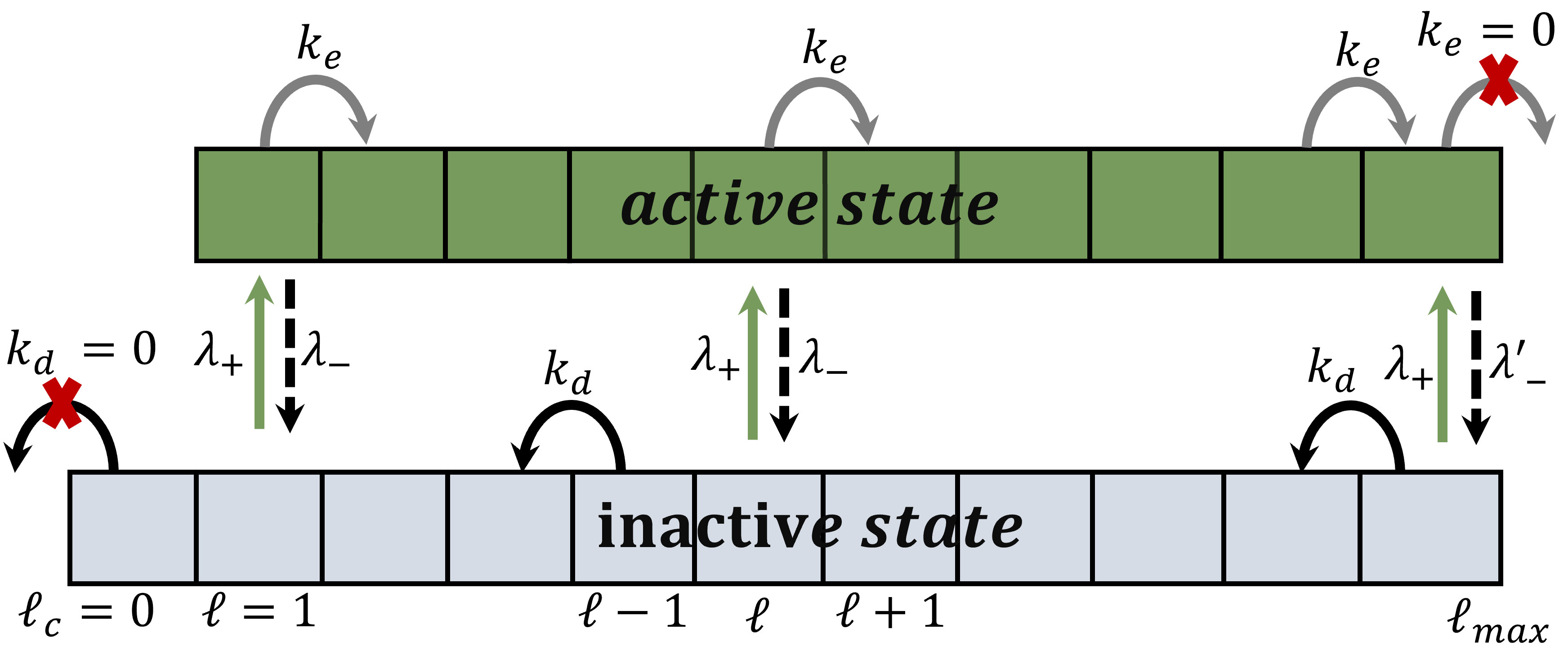}
	\caption{Schematic description of the elongation-shortening-switching kinetics of a polymeric filament in the stochastic model (see text for details).}
	\label{fig1}
\end{figure}

%%%%%%%%%%%%%%%%%%%%
In this model the filament length $\ell$ is treated as a discrete variable $\ell=0,1,…,\mathcal{N}$. Equivalently, the distal tip of the filament is represented by a hypothetical particle at the discrete position ${\ell}$ ($\ell=0,1,…,\mathcal{N}$) on a one-dimensional lattice. The filament can be in one of the two activity states: an inactive state $(i)$, in which the filament can only shorten, and an active state $(a)$, in which filament can only elongate. Thus, the elongation-shortening kinetics of the filament can be, equivalently, viewed as a correlated random walk of the hypothetical particle coinciding with the distal tip of the filament; in the state $i$ the particle can step only towards $\ell=0$ whereas the state $a$ it can step only away from the $\ell=0$.
In the state $i$, the filament shortens by one unit with rate (i.e., probability per unit time) $k_d$. In the state $a$, the filament elongates by one unit with rate $k_e$. Transitions between the two activity states occur at fixed rates (probabilities per unit time): activation $(i\rightarrow a)$ with rate $\lambda_{+}$ and deactivation ($a \rightarrow i$) with rate $\lambda_{-}$ (see Fig.\ref{fig1}). Strictly speaking, each of the rates of transitions between the states of the generic model represent the probability of the corresponding transitions per unit time where the stochastic kinetics is assumed to be Markovian. In the example of microtubule, $\lambda_{+}$ and  $\lambda_{-}$ can be related to rescue and catastrophe rates; however, factors such as the evolving GTP cap, lattice age, GTP-rich islands, or hydrolysis history are not explicitly captured in this generic model. When applied to the similar kinetics of other filaments, the microscopic interpretation of these same rates may differ. The kinetics is governed by a set of coupled forward master equations for the probabilities $P^i({\ell},t)$ and $P^a({\ell},t)$:

%%%%%%%%%%%%%%%%%%%%%%%%%%%%%
\begin{equation}
	\frac{dP^a(\ell,t)}{dt} = k_e P^a(\ell-1,t) - (k_e + \lambda_{-})P^a(\ell,t) + \lambda_{+} P^i(\ell,t) \\
	\label{eq1-model}
\end{equation}
\begin{equation}
	\frac{dP^i(\ell,t)}{dt} = k_d P^i(\ell+1,t) - (k_d + \lambda_{+})P^i(\ell,t) + \lambda_{-} P^a(\ell,t)
	\label{eq2-model}
\end{equation}
%%%%%%%%%%%%%%%%%%%%%%%%%%%%%%
for $1 < {\ell} < \mathcal{N}$.

The special value $\ell=0$ represents the vanishing length of a MT or the critical length of the telomere where the cell enters senescence irreversibly. This absorbing state state at ${\ell}=0$ can be reached only from ${\ell}=1$ in the inactive state; so the corresponding master equation is given by
\begin{equation}
	\frac{dP^i(0,t)}{dt} = k_d P^i(1,t) 
	\label{eq7-model}
\end{equation}
So far as the dynamics at $\ell=1$ is concerned, the master equations are given by
\begin{equation}
	\frac{dP^a(1,t)}{dt} =\lambda_{+} P^i(1,t) - (k_e + \lambda_{-})P^a(1,t)  
	\label{eq5-model}
\end{equation}
\begin{equation}
	\frac{dP^i(1,t)}{dt} = k_d P^i(2,t) + \lambda_{-} P^a(1,t) - (k_d+\lambda_{+}) P^i(1,t) 
	\label{eq6-model}
\end{equation}

Elongation of the filament beyond the maximum length $\ell=\mathcal{N}$ is not allowed. To ensure probability flux conservation at this boundary the rate $\lambda'_{-}$ is replaced by an effective rate $\lambda'_{-}= \lambda_{-}+k_e$. This is a stochastic hard-cap prescription; it does not explicitly describe the mechanics of the filament-boundary contact like, for example, elastic force generation, and bending, or buckling of the filament. The corresponding master equations are given by
\begin{equation}
	\frac{dP^a(\mathcal{N},t)}{dt} = k_e P^a(\mathcal{N}-1,t) + \lambda_{+} P^i(\mathcal{N},t) - \lambda'_{-}P^a(\mathcal{N},t)
	\label{eq3-model}
\end{equation}
\begin{equation}
	\frac{dP^i(\mathcal{N},t)}{dt} = \lambda'_{-} P^a(\mathcal{N},t) - (k_d + \lambda_{+})P^i(\mathcal{N},t). 
	\label{eq4-model}
\end{equation}
We introduce a dimensionless parameter $R=(k_e\lambda_{+})/(k_d\lambda_{-})$ which is a ratio of the net tendency toward elongation following activation and the competing tendency of shortening upon deactivation. This single parameter serves to classify different dynamical regimes:  $R<1$ (and $R > 1$) correspond to two regimes dominated by filament length shortening (elongating), whereas $R=1$ corresponds to an effective drift free point which, as we argue later, is a critical point.\\

%%%%%%%%%%%%%%%%%%%%%%
%%%%%%%%%%%%%%%%%%%%%%%%
\section{Mean First Passage Time (MFPT) for unconstrained filament length}
 Before imposing a boundary at a finite ${\ell}=\mathcal{N}$, we consider the simpler special case where unconstrailed growth of the filament to any length is allowed. 
We now address the key question: how long it would take, on the average, to reach ${\ell}=0$? Since, by definition, ${\ell}=0$ is an absorbing state, the answer to the above question requires essentially calculation 
of a mean first-passage time (MFPT). Using the formalism of backward master equation, combined with the survival probability, the exact expression for the MFPT of the filament for the initially $inactive$ and initially $active$ states are given by.
\begin{equation}
		T^{\,i}(\ell)=\mu\,\ell,\qquad
		T^{\,a}(\ell)= \mu \biggl[\ell + \frac{k_d + k_e}{\lambda_{-} + \lambda_{+}} \biggr]
	\label{MFPTsM}
\end{equation}
where
\begin{equation}
	\mu = \frac{\lambda_{-}+\lambda_{+}}{\,k_{d}\lambda_{-}-k_{e}\lambda_{+}},  
\end{equation}
A heuristic derivation of (\ref{MFPTsM}) provides a deeper physical understanding. Beginning in an {\it inactive state} the filament spends an average time $(\lambda_{+} +\lambda_{-})^{-1}$ in one cycle during which the average net change in length is $(k_d\lambda_{-}-k_e\lambda_{+})$ (the length of each unit is one, in our notation). Therefore, the mean drift velocity is
\begin{equation} 
V_{drift}  = \frac{\,k_{d}\lambda_{-}-k_{e}\lambda_{+}}{\lambda_{-}+\lambda_{+}} = \frac{1}{\mu}
\end{equation}
and $T^{\,i}(\ell) = {\ell}/V_{drift} = \mu {\ell}$, as in ({\ref{MFPTsM}). In contrast, if the filament is initially in an {\it active state}, before transitioning to the inactive state
 it dwells, on the average, an extra time $1/\lambda_{-}$ during which it moves an average distance $k_{e}/\lambda_{-}$. In order to reach the absorbing state at ${\ell}_c=0$, the tip of the filament has to retrace the same extra distance with its average speed $V_{drift} = (k_{d}\lambda_{-}-k_{e}\lambda_{+})/(\lambda_{-}+\lambda_{+})$. Thus, if the initial state is {\it active}, the total mean time required to reach the absorbing state at ${\ell}_c=0$ is 
 \begin{eqnarray} 
 T^{\,a}(\ell) = T^{\,i}(\ell) + \frac{1}{\lambda_{-}} + \biggl(\frac{k_e}{\lambda_{-}}\biggr) \biggl(\frac{1}{V_{drift}}\biggr) 
 \end{eqnarray}
which is identical to the expression for $T^{\,a}(\ell)$ in (\ref{MFPTsM}). 

Note that both $T^{\,i}(\ell)$ and $T^{\,a}(\ell)$ are linear in ${\ell}$. The  dimensionless ratio 
$(k_d + k_e)/(\lambda_{-} + \lambda_{+})$ reflects the relative rates of competing processes of elongation-shortening of the filament and activation-inactivation process. 
Moreover, from the exact analytical expressions for MFPTs we find that as $\lambda_{+}+\lambda_{-} \to \infty$, keeping $k_{d}+k_{e}$ fixed at a finite non-zero value  $T^{\,a}(\ell) \to T^{\,i}(\ell)$. Both this implications of the expressions are consistent with our observations on the plots in Fig.\ref{fig2} where numerical values obtained from analytical formulae (\ref{MFPTsM}) for three sets of values of the rate constants are compared with the corresponding numerical data obtained from our Monte Carlo simulations of the same system. In those Figs. we choose $V_{drift} = (k_d\lambda_{-}-k_e\lambda_{+})>0$, i.e., there is an overall bias towards the absorbing state ${\ell}_c=0$; a stronger bias (i.e., higher value of $V_{drift}$) results in a shorter MFPTs.

%%%%%%%%%%%%%%%%%%%%%%
\begin{figure}[t]
	\centering
	\includegraphics[angle=0,width=0.9\linewidth]{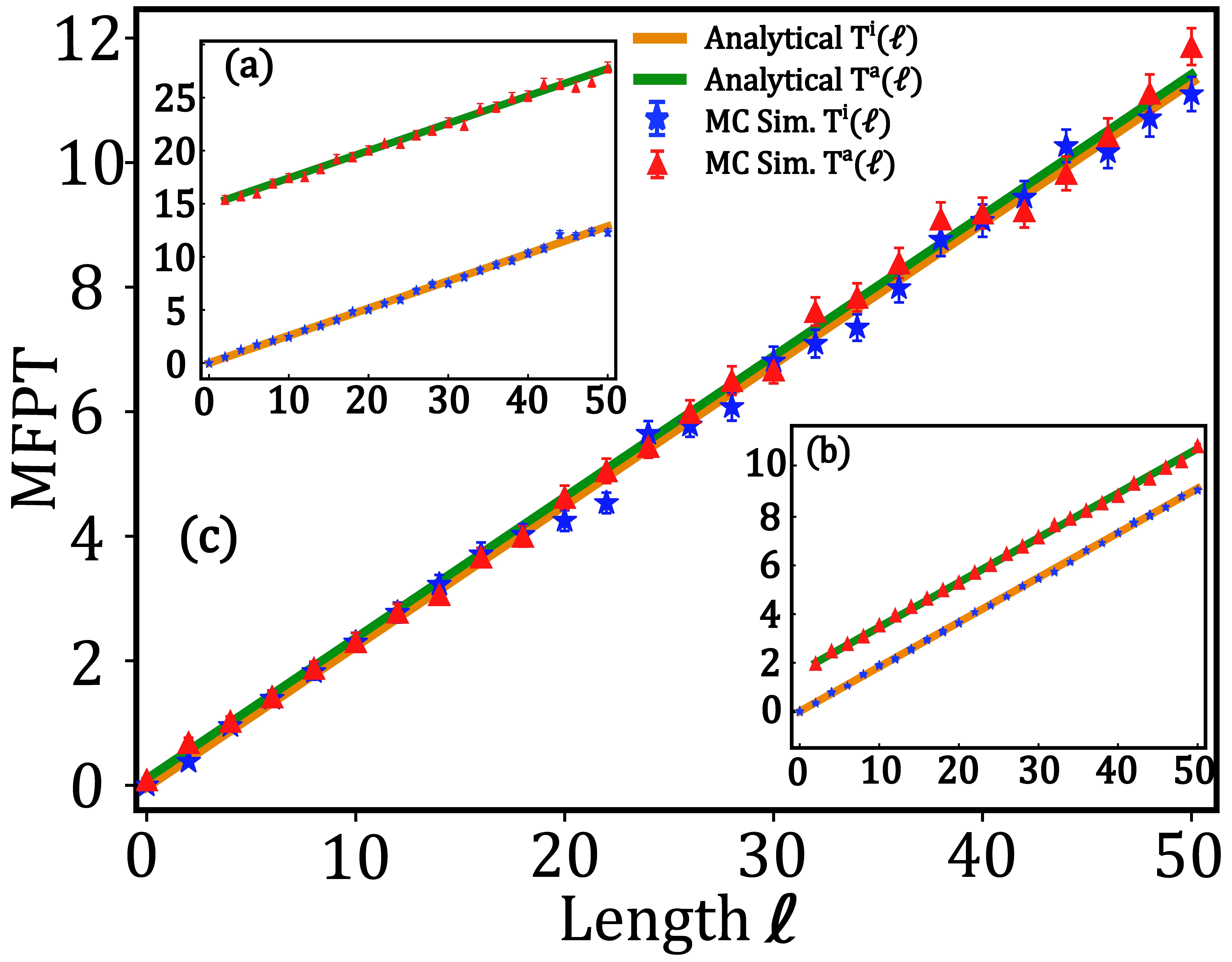}
	\caption{(Color online)
		Mean first-passage time (MFPT) versus length $\ell$ for three representative parameter sets, shown (a)–(c). 
		The continuous lines show the analytical results, and the symbols show the corresponding Monte Carlo simulation data. 
		For each case, the two continuous lines correspond to the MFPTs $T^{i}(\ell)$ and $T^{a}(\ell)$ associated with the two initial states. 
		The parameters are: (a) $k_d=50,\; k_e=45,\; \lambda_{+}=0.8,\; \lambda_{-}=0.9$; 
		(b) $k_d=50,\; k_e=48,\; \lambda_{+}=5,\; \lambda_{-}=6$; 
		(c) $k_d=50,\; k_e=48,\; \lambda_{+}=100,\; \lambda_{-}=115$.}
	\label{fig2}
\end{figure}
%%%%%%%%%%%%%%%%%%%%%%%%

\begin{figure*}[t]
	\centering
	\includegraphics[width=\textwidth]{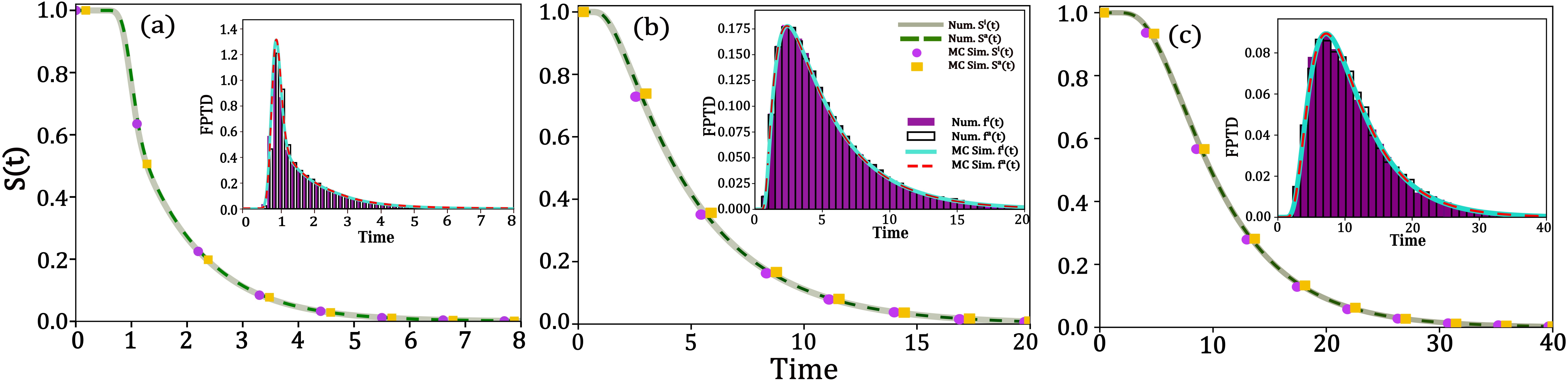}
    \caption{(Color online) FPT distributions and survival probabilities.
    		By keeping fixed $k_e=45$ and $k_d=50$, respectively, in all three panels, while the switching rates are varied
    		as follows: (a) $\lambda_{+}=1,\;\lambda_{-}=2$;
    		(b) $\lambda_{+}=8,\;\lambda_{-}=9$; and
    		(c) $\lambda_{+}=100,\;\lambda_{-}=105$.
    		The numerical results are verified by Monte Carlo simulations.
    	 Together, $S^{\sigma}(t)$ and $f^{\sigma}(t)=-\partial_tS^{\sigma}(t)$ provide the complete time-resolved absorption statistics, including long-lived trajectories that cannot be resolved from the MFPT alone. As switching rates increase from panel (a) to panel (c), \(R\) approaches unity from below. This weakens the net shortening drift; consequently, the peak of the FPTD shifts towards later times, and its long-time tail becomes broader.}
	\label{fig3}
\end{figure*}
The variance of the distributions of the FPTs for the $inactive$ and $active$ states are
\begin{equation}
	Var[\tau^i(\ell)] = \langle (T^{i}({\ell}))^2 \rangle -  \langle (T^{i}({\ell})) \rangle^2  = (2\nu -\mu^2)\ell
	\label{Varience I}
\end{equation}
\begin{equation}
	Var[\tau^a] = \ell(2\nu-\mu^2) + [ \frac{2k_d(\nu-\mu^2)}{\lambda_{+}} +  \frac{2k_d\mu -k_d^2\mu^2-1}{\lambda^2_{+}}]
	\label{Varience A}
\end{equation}
where
\begin{equation}
 \nu = \frac{-2 -2\mu(k_e-k_d-\lambda_- - \lambda_+)+2k_e(k_d+\lambda_+)\mu^2}{k_e\lambda_+-k_d\lambda_-}
\end{equation}

In the limit $R \to 1$, $V_{drift} = (k_d\lambda_{-}-k_e\lambda_{+}) \to 0$, i.e., the kinetics of the distal tip of the polymer becomes purely diffusive. At this point
$Var[\tau^i(\ell)] \to \infty$ and $Var[\tau^a] \to \infty$, i.e., the fluctuations in the FPT diverges while, simultaneously, ($T^{\sigma}(\ell) \to \infty$) for both $\sigma \equiv i,a$ which 
expresses the phenomenon of `critical slowing down'  at the critical point $R=1$.\\

%%%%%%%%%%%%%%%%%%
\section{Mean First Passage Time (MFPT) for constrained maximum filament length}
%%%%%%%%%%%%%%%%%%

Next, we assume a reflecting boundary at $\ell=\mathcal{N}$; at this boundary the probability flux cannot leave the domain, it is entirely reflected back.
\begin{equation}
	\,T^{\,i}(\ell)= \mu\,\ell
		+\Big[\frac{1+(\lambda_+/\lambda'_-)}{k_d}-\mu\Big]\, \biggl(\frac{1}{\,r_0^{\mathcal{N}-1}}\biggr) \biggl[\frac{(r_0^\ell-1)}{(r_0-1)\,}\biggr]\,
	\label{MFPT WB1}
\end{equation} 
where $\lambda'_{-} = \lambda_{-} + k_{e}$ and 
\begin{equation}
r_0= \frac{k_d(k_e+\lambda_{-})}{k_e(k_d+\lambda_{+})} =\frac{k_d/(k_d+\lambda_{+})}{k_{e}/(k_e+\lambda_{-})};
\end{equation} 
the last step shows that $r_{0}$ is not just a dimensionless parameter; rather, it measures the ratio of the effective shortening tendency and the effective elongation tendency. 
Since in the decay-dominated regime ($k_d\lambda_{-}>k_e\lambda_{+})$), the coefficient 
$\Big[\frac{1+(\lambda_+/\lambda'_-)}{k_d}-\mu\Big] < 0$, the MFPT $T^{i}(\ell)$ in the presence of a reflecting boundary at $\mathcal{N}$ is smaller than that in the absence of that boundary; physical origin of these shorter MFPT is that the reflecting boundary prohibits time-consuming excursions beyond $\mathcal{N}$ that would otherwise be possible in the open system.
Similarly, 
 \begin{equation}
	\,T^{\,a}(\ell)=\frac{(k_d+\lambda_+)T^{\,i}(\ell)-k_d\,T^{\,i}(\ell-1)-1}{\lambda_+}\,
	\label{MFPT WB2}
\end{equation}

 In the decay-dominate regime $k_d\lambda_{-}>k_e\lambda_{+}$ and, therefore, not only $V_{drift} > 0$, we also have $r_{0} > 1$. Consequently, we have not only $(r_0^\ell-1)>0$, but also $r^{-(\mathcal{N}-1)}_0 \to 0$ as $\mathcal{N}\rightarrow \infty$, i.e., as the boundary moves to $\infty$, and hence, $T^{\,i}(\ell) \to \mu\,\ell$, as it must. The survival probability $S(t)$ decays monotonically unity to zero. The active and inactive curve overlaps, indicating the constraint strongly suppresses the memory of the internal state. Physically, the constraint removes the persistence growth pathway of active state and funnels both initial condition into nearly the same route towards absorption. From Fig.\ref{fig3} (a)-(c), decay become progressively broader and shifts to longer times consistence with FPTDs. \\

At the drift-balance condition $R=1$, the average drift becomes zero, and the two characteristic roots merge to become identical. Therefore, Eq.~(\ref{MFPT WB1}) must be solved separately for this condition. For a finite upper boundary $\mathcal{N}$, we obtain
	\begin{equation}
		T^{\,i}(\ell)
		=
		v_2\ell^2+
		\left[\Delta-v_2(2\mathcal{N}-1)\right]\ell,
		\label{MFPT-critical}
	\end{equation}
	where
	$v_2=-(\lambda_{-}+\lambda_{+})/
	(2k_dk_e+k_e\lambda_{+}+k_d\lambda_{-})<0$
	and
	$\Delta=[1+\lambda_{+}/\lambda'_{-}]/k_d$.
	If the initial length $\ell$ is kept fixed, then
	$T^{\,i}(\ell)\sim-2v_2\ell\mathcal{N}\rightarrow\infty$
	as $\mathcal{N}\rightarrow\infty$. However, if
	$\ell/\mathcal{N}$ is kept fixed, then
	$T^{\,i}\propto\mathcal{N}^2$.
	Thus, a finite upper boundary renders the critical MFPT finite, whereas the MFPT diverges in the open-system limit. The active-state MFPT also follows the same finite-size scaling. Therefore, at criticality, the upper boundary is not a perturbative correction; it is the finite-size regularizer that controls how the divergent open-system lifetime is approached.

For $R < 1$, the process can be treated as essentially a drift-diffusion process for which the FPTD is known to have the form \cite{Redner2001}
\begin{equation} 
F({\ell},t) = \frac{{\ell}}{\sqrt{4 \pi D t^3}} exp\biggl[- \frac{(\ell+vt)^2}{4Dt} \biggr], 
\label{eq-FPTDa}
\end{equation}
where $\ell$ is the distance from the absorbing boundary, $D$ is the diffusion constant and $v$ is the drift velocity. This expression can be re-expressed as 
\begin{equation} 
F({\ell},t) \propto t^{-3/2} exp(- A/t) exp(- B t), 
\label{eq-FPTDb}
\end{equation}
with $A = \ell^2/(4Dt)$ and $B = v^2 t/(4D)$. In the limit of large enough $t$, the FPTD is well approximated as $F({\ell},t) \propto t^{-3/2} exp(- B t)$, which is 
essentially the asymptotic form reported in \cite{bicout97}. Our FPTDs simulation data (see insets of Fig.\ref{fig3}) are well described by Eq.(\ref{eq-FPTDb}).

\section{Summary}
In this work we have studied the stochastic kinetics of elongation, shortening, and switching of a polymeric filament using a generic model. Our main focus is on the first-passage time (FPT) of the filament, which is the time required for the distal tip to reach the absorbing state at $\ell=0$ for the first time, starting from an initial length $\ell>0$. By employing backward master equations, we derived analytical expressions for the mean and variance of the first-passage time. Additionally, from the equations formulated above, we have numerically evaluated the first-passage time distributions and survival probabilities. By comparing these numerical results with our corresponding Monte Carlo simulation data we have established the validity of the theoretical formalism even in regimes where FPT cannot be obtained analytically.The dimensionless parameter $R$ quantifies the relative strength of the effective elongation and shortening tendencies. Previous studies have noted that $R=1$ separates shortening-dominated from elongation-dominated filament growth \(\cite{dogterom93,mishra05}\). Here we have established its significance as a critical point for first-passage behavior and explored how this balance is affected by finite geometry. The exact solution for finite $\mathcal{N}$ provides a unified description of three distinct return mechanisms. For $R<1$, absorption is primarily influenced by the bulk shortening bias, and the effects of a distant upper boundary gradually diminish. At $R=1$, confinement limits drift-free excursions and transforms the infinite-mean open process into a finite, system-size-controlled lifetime. For $R>1$, the presence of the finite cap eliminates the open escape pathway and replaces potential nonreturns with rare returns against the elongation bias. Thus, the upper boundary not only affects the numerical value of the MFPT but also alters the statistical characteristics and physical mechanisms of the return process.The main scientific advancement presented here is a unified analytical classification of how the kinetic control parameter $R$ and the geometric scale $\mathcal{N}$ together determine filament lifetime. Due to its general formulation, our model serves as a controlled baseline for future studies that may incorporate factors such as chemical memory, age-dependent switching, spatially varying rates, or additional internal states. \\

\section{Acknowledgment}
 AG and MS gratefully acknowledge the LNM Institute of Information Technology, Jaipur, for its institutional support. MS also gratefully acknowledges the Institute for providing financial assistance through the Institute Fellowship.

%%%%%%%%%%%%%%%%%%%%%%%%%%%%%%%%%%%%%%%%%%%%%%%%%%%%%%%%%%%%%
%  References
%%%%%%%%%%%%%%%%%%%%%%%%%%%%%%%%%%%%%%%%%%%%%%%%%%%%%%%%%%%%%
%\clearpage

\end{document}